\documentclass[12pt]{article}
\usepackage{graphicx}
\usepackage{gensymb}
\usepackage[colorlinks=true,
            linkcolor=blue,
            urlcolor=blue,
            citecolor=blue]{hyperref}
\usepackage{url}
\usepackage
[        a4paper,
        hmargin=1.135in
]{geometry}

\newcommand\pubdate{\today}
\newcommand{\three}{3$\times$1$\times$1$\,\rm m^3$}

\def\UCL{Department of Physics and Astronomy\\
University College London, London WC1E 6BT, United Kingdom}

\def\Title#1{\begin{center} {\Large #1 } \end{center}}
\def\Author#1{\begin{center}{ \sc #1} \end{center}}
\def\Address#1{\begin{center}{ \it #1} \end{center}}

\newcommand\pubblock{\rightline{\begin{tabular}{l} 
         \pubdate  \end{tabular}}}
\newenvironment{Abstract}{\begin{quotation}  }{\end{quotation}}
\newenvironment{Presented}{\begin{quotation} \begin{center} 
             PRESENTED AT\end{center}\bigskip 
      \begin{center}\begin{large}}{\end{large}\end{center} \end{quotation}}

\begin{document}
\begin{titlepage}
\pubblock

\vfill
\Title{ProtoDUNEs}
\vfill
\Author{Laura Manenti}
\Address{\UCL}
\vfill
\begin{Abstract}
These proceedings review the two DUNE prototype detectors, namely Single- and Dual-Phase ProtoDUNEs. The detectors, both employing liquid argon Time Projection Chambers (LAr TPCs), are currently being built at CERN as part of the ProtoDUNE experimental programme. Such R\&D programme aims at validating the prototypes design and technology, which will eventually be applied to the DUNE Far Detector at the Sanford Underground Research Facility (SURF). 
\end{Abstract}
\vfill
\begin{Presented}
NuPhys2016, Prospects in Neutrino Physics\\
Barbican Centre, London, UK,  December 12--14, 2016
\end{Presented}
\vfill
\end{titlepage}
\def\thefootnote{\fnsymbol{footnote}}
\setcounter{footnote}{0}

\section{Introduction}
The Deep Underground Neutrino Experiment (DUNE), which mainly aims at studying long-baseline neutrino oscillations, foresees to employ 40,000 tonnes of liquid argon (LAr) split among four detectors to be installed at the Sanford Underground Research Laboratory in Lead,  South Dakota. These will be used as far detectors for neutrinos produced 1,300 km away at the Fermi National Accelerator Laboratory in Batavia, Illinois.

As the task is not trivial, to gain experience in building and operating such large scale LAr detectors, an R\&D programme is currently underway at CERN. Such programme will operate two prototypes in a dedicated test beam by 2018, with the specific aim of testing the prototypes design, assembly and installation procedures, the detectors operations, as well as data acquisition, storage, processing and analysis under beam conditions.

The two prototypes will both employ Liquid Argon Time Projection Chambers (LAr TPCs) as detection technology, with one prototype only using liquid argon, hereby called ProtoDUNE Single-Phase (SP), and the other using argon in both its gaseous and liquid state, thus the name ProtoDUNE Dual-Phase (DP). 

In the past few years, there has been an extensive and increasing effort in terms of R\&D, funding, and manpower towards LAr TPCs for studies of neutrinos. These detectors are  in fact very close to what a ``perfect'' neutrino detector is required to do: firstly, providing a large detection mass (at a relatively low cost) to compensate for the small neutrino cross-section; secondly, granting the possibility of distinguishing between $\nu_\mu$ and $\nu_e$ signals. 

The first ever built tonne-scale LAr TPC was ICARUS-T600~\cite{ICARUS}, which was located at Gran Sasso and employed 760 tonnes of liquid argon ($\sim$460 tonnes active mass). The experiment operated between 2010 and 2013, collecting data using the CERN Neutrinos to Gran Sasso (CNGS) beam. It has now been refurbished at CERN and is currently being moved to Fermilab, where it will be used as the far detector in the Short-Baseline Neutrino (SBN) programme.
Since then many other SP LAr TPCs have been built, such as ArgoNeuT, LArIAT, MicroBooNE, and the 35 Ton Prototype.
As for the dual-phase technology, a 3 litres chamber and a 250 litres one have been constructed and operated at CERN by ETHZ. 
Currently, the WA105 \three\ prototype, a 5 tonnes active dual-phase LAr TPC, is being commissioned at CERN~\cite{wa105}.
The \three\ demonstrator will take data with cosmic muons and, as it features many of the same technical challenges of the ProtoDUNE-DP, it will serve the purpose of test-bench for the detector to come. 

\section{ProtoDUNEs overview}
Both protoDUNEs will be located in a dedicated building in the North Area at CERN (EHN1) and placed along two separate beam lines (H2 and H4), provided by the CERN Super Proton Synchrotron (SPS). 
Both experiments are foreseen to start taking data in autumn 2018 till the planned SPS beam shutdown the same year. 

Both detectors will have similar sizes and active mass (450\,t for the ProtoDUNE-SP and 300\,t for the ProtoDUNE-DP). The cryostats, which contain the LAr target and the TPCs, are identical, with outer dimensions of roughly 11$\times$11$\times$11\,m$^3$ and a total capacity of around $\sim$700\,t of LAr. 
Two independent cryogenic systems fill the tanks and recirculate the liquid argon to guarantee a purity level of around 100\,ppt. This is to ensure that the electrons produced by the ionising particles in the medium do not get trapped and reach the charge readout system to later enable track reconstruction. 

\subsection{Cryostat technology}
The cryostats, which are currently being constructed, make use of a ``membrane'' technology developed by commercial company GTT/France to store and ship liquified natural gas (LNG) at a temperature of $-163$\,\degree C.
The cold vessel is housed in a warm support structure and comprises a primary corrugated stainless steel membrane and a secondary membrane made of Triplex (a composite laminated material composed of a thin sheet of aluminum between two layers of glass cloth and resin~\cite{gtt}). 
The secondary membrane is inserted between two insulation layers made of polyurethane, with the first layer directly touching and supporting the primary membrane~\cite{gtt}.
The same company also built the \three\ cryostat, albeit using a slightly different design with only one insulated membrane.

A similar membrane cryostat was firstly tested in 2013 with the 35 Ton Prototype. Such membrane cryostat was designed by Fermilab engineers in collaboration with IHI Corporation of Tokyo, Japan.
The main goal of the prototype (initially not equipped with a TPC) was to demonstrate that a non-evacuable membrane cryostat of this dimensions (i.e. 35 tonnes of LAr) could reach an oxygen contamination below 200\,ppt~\cite{35t, Montanari:2014, Montanari:2015}.
The cryostat was successfully operated and after the 11$\rm ^{th}$ exchange of volume an electron drift time of 3\,ms had been achieved~\cite{Tope2014}. This corresponds to an oxygen concentration of 100\,ppt w/V according to the formula~\cite{Bakale1976, EmissionDetectors, Buckley1989, Swan1963}:
\begin{equation}
\tau [\rm \mu s]= \frac{0.3\, \rm ppm\,w/V \cdot \mu s}{\rho_{\rm O_2} [\textrm{parts of }\rm O_2\;w/V]}
\end{equation}
where $\rho_{\rm O_2}$ is the oxygen concentration expressed in units of $\textrm{parts of }\rm O_2$ weight by volume (w/V).

\section{ProtoDUNE Single-Phase}

\begin{figure}[t]
	\begin{center}
	\includegraphics[width=1\textwidth, trim={0cm 1.2cm 0cm 3cm}, clip=true]{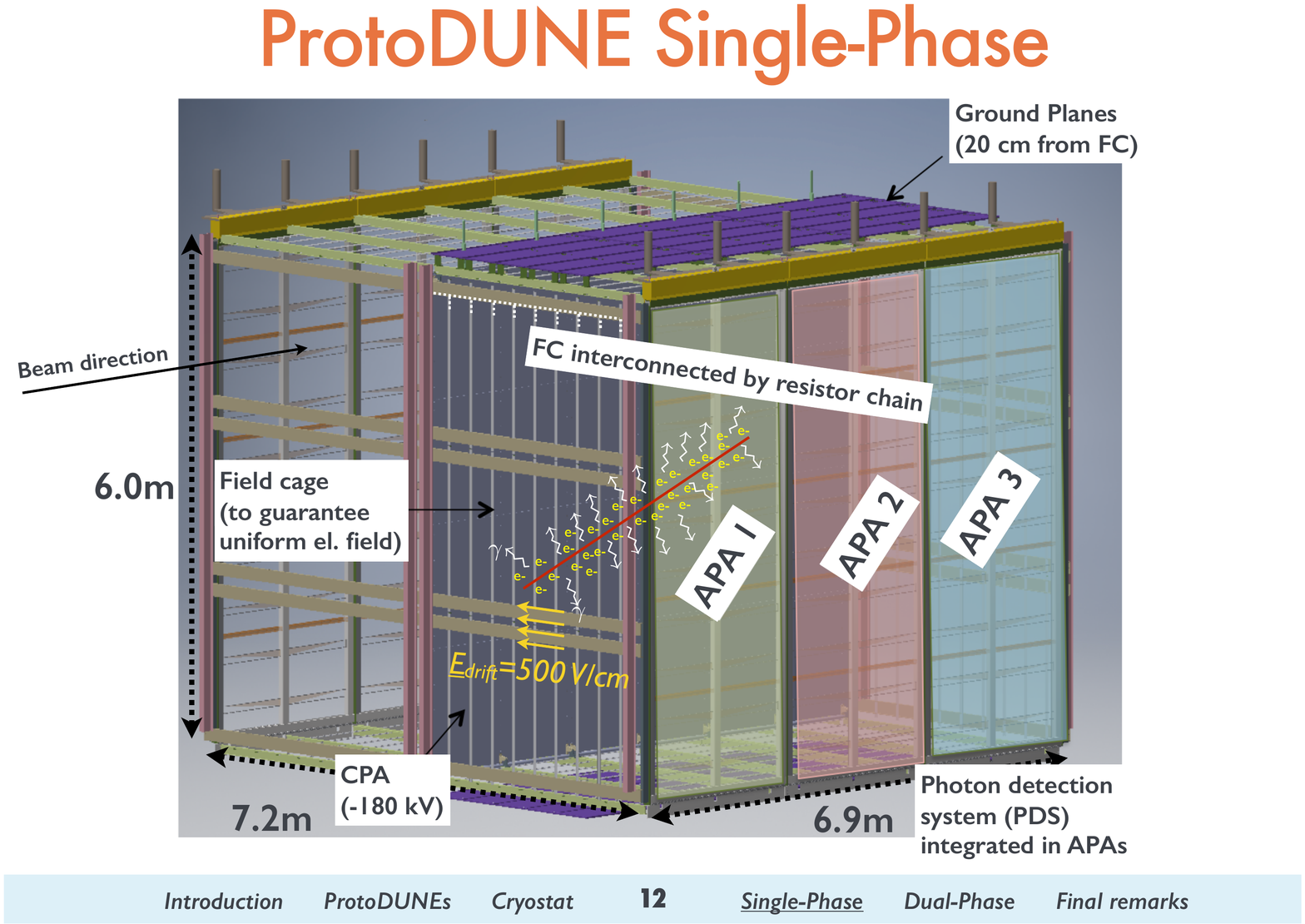}	
	\caption{Design of the ProtoDUNE-SP detector. A description of the main parts in given on the picture itself.}
	\label{fig:CAD_SP}
	\end{center}
\end{figure}

Before proceeding with the description of the ProtoDUNE-SP, let us see how a single-phase LAr TPC works in general.
Ionising particles that traverse the liquid argon will produce electrons and scintillation photons along the track. The charge and light signals arising from the aforementioned processes are together used to make a 3D reconstruction of the particles' track in the following way. The electrons produced along the track drift under the electric field between cathode and anode, where they induce a transient signal on two wire planes, called induction planes, and are collected on a third one, enabling a 2D image of the track. The scintillation photons are instead collected by photodetectors (usually PMTs) and provide the trigger signal and the reference time for the interaction. By knowing the drift velocity at a given electric field and combining the time zero of the event with the distance travelled by the electrons, the depth of the ionisation track can be reconstructed, hereby achieving a full 3D image of the event. 

We now may look at how this is achieved in the the ProtoDUNE-SP. Figure~\ref{fig:CAD_SP}
shows a CAD design of the detector inside the cryostat. It is composed of one single cathode plane at the centre and two anode planes on the sides. 
The high-voltage provided to the cathode is $-180$\,kV, such that an electric field of 500\,V/cm is present between the cathode and each anode. 
The particles coming from the test beam traverse the detector along the cathode plane. 
A field cage interconnected by a resistor chain guarantees the uniformity of the field. 
Ground planes are placed on the top and bottom of the field cage (purple in the figure) to prevent the electric field from entering the ullage region in the top (where the high-voltage feedthrough is inserted) and to shield the cryogenic pipes in the bottom on the membrane. 
Both anode planes comprise three adjacent Anode Plane Assemblies (APAs).
Each APA consists of a grid, two induction planes oriented at $\pm$37.5\,\degree\ with respect to the vertical axis, and a collection plane. Behind the collection plane another layer of wires, a grounded mesh, shields the photon detection system behind (e.g. from ``ghost'' tracks traversing the collection plane)~\cite{TDR_SP}. 

As seen previously, a particle entering the LAr will ionise the medium, leaving electrons around the track which will move from the cathode towards one of the two anodes under the influence of the electric field. As just mentioned, the induction and collection planes are screened by a grid. The reason for such grid is the following. Positive ions moving to the cathode induce on the anode plane a signal of the same polarity as that from electrons moving to the anode. It was shown that this induction effect can be eliminated by inserting a grid in front of the induction/collection wire planes and biasing it appropriately~\cite{Bunemann1949}.
The inductions planes are transparent to the charges and when the drifting electrons pass through them they induce a bipolar signal. In principle, for the event 3D reconstruction only one induction and collection plane are needed, but an extra one allows to reconstruct tracks with ambiguous orientation. 

As for the photon detection system, the ProtoDUNE-SP uses Silicon Photomultipliers (SiPMs). Light guides are embedded in the APA structure. These guides are coated in Tetraphenyl-Butadiene (TPB), a wavelength shifter which allows the VUV LAr scintillation light (the spectrum being peaked at $\sim$128\,nm) to be shifted into the visible range. Once wavelength-shifted, photons are collected from the SiPMs mounted at the end of the light guide. 

\section{ProtoDUNE Dual-Phase}

\begin{figure}[t]
	\begin{center}
	\includegraphics[width=1\textwidth, trim={0cm 1.2cm 0cm 1.5cm}, clip=true]{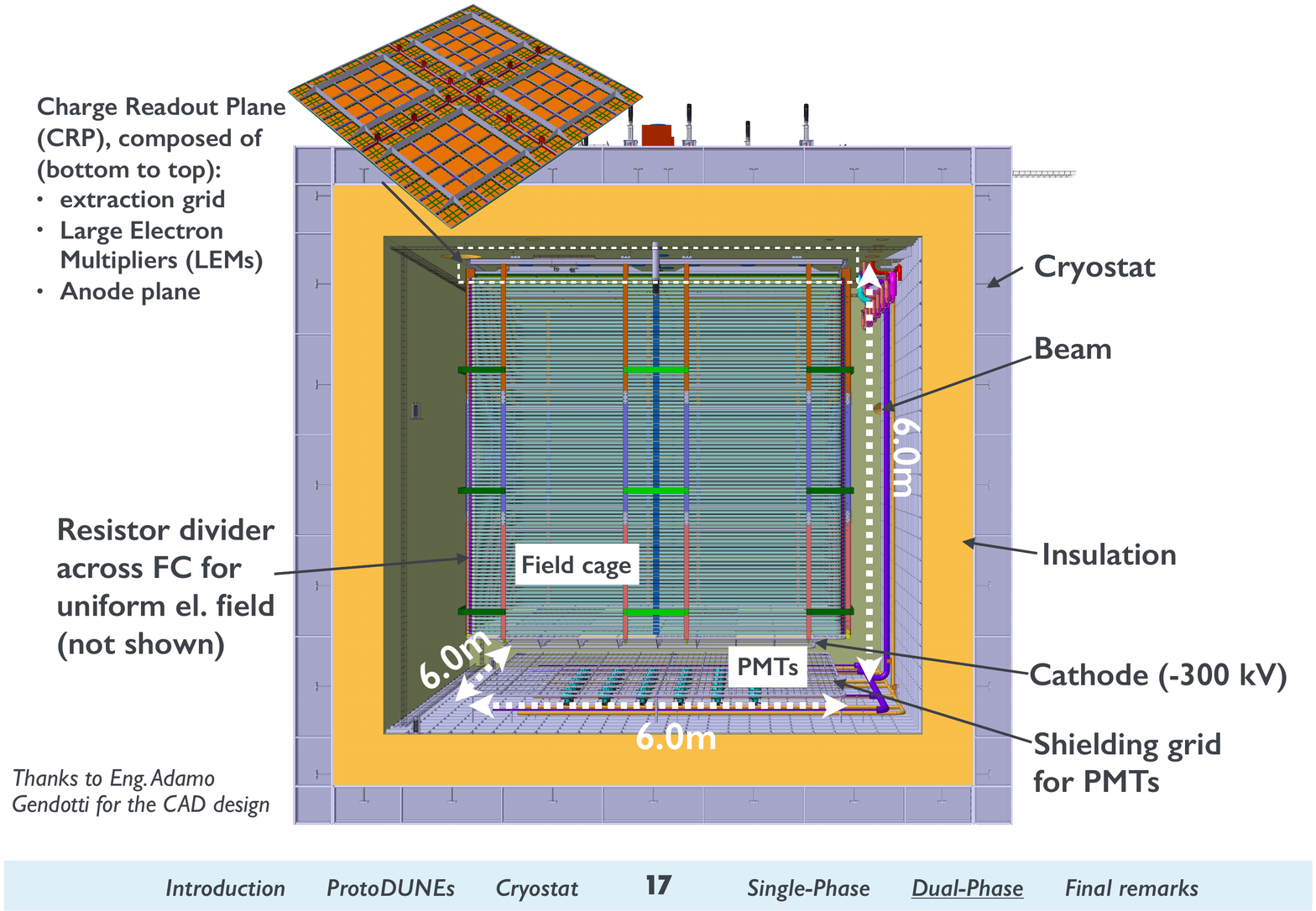}	
	\caption{Design of the ProtoDUNE-DP detector with main parts highlighted.}
	\label{fig:CAD_DP}
	\end{center}
\end{figure}

To improve image resolution and increase the signal to noise ratio, dual-phase technology may be used. Let us see how this is achieved by directly describing the ProtoDUNE-DP. 

Figure \ref{fig:CAD_DP} shows the CAD design of the detector housed in the cryogenic vessel. The TPC comprises two independent structures: the drift cage and the charge readout plane (CRP). The first structure is composed of a ground grid with PMTs underneath, a cathode, and the drift cage, whose aluminum shaping field rings allow a uniform electric field. The second structure consists of an extraction grid, Large Electron Multipliers (LEMs), and the anode readout plane.

Liquid argon fills the cryostat up to 5\,mm above the extraction grid, with the rest of the tank being filled with gaseous argon. The scintillation photons produced by particles traversing the LAr (note the beam entrance on the right in Fig.~\ref{fig:CAD_DP}) are detected by TPB coated PMTs. As for the ProtoDUNE-SP, the light signal provides the time reference of the event (note that the test beam itself may also give the time reference). The ionisation electrons along the particle's track are shifted upwards under the 500\,V/cm electric field in the drift cage. 
For 6\,m drift distance, 300\,kV must be provided to the cathode. As of today, the high-voltage feedthrough used in the \three\ detector has been successfully tested in liquid argon down to $-300$\,kV~\cite{Cantini2016}. The same power supply and an analogous design will be used for the high-voltage feedthrough for the ProtoDUNE-DP. 

\begin{figure}[t]
	\begin{center}
	\includegraphics[width=0.65\textwidth, trim={0cm 0cm 0cm 0cm}, clip=true]{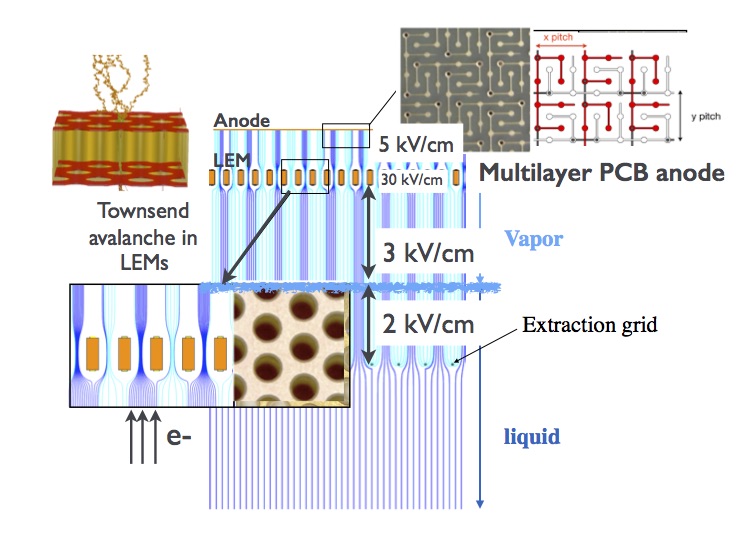}	
	\caption{Schematic of the charge readout plane: the electrons are extracted from the liquid into the gas, where they are collimated in the LEMs, undergoing a process called ``Townsend avalanche''. The amplified charge is then read by the multilayer PCB anode.}
	\label{fig:LEM}
	\end{center}
\end{figure}

As shown in Fig.~\ref{fig:LEM}, the electrons drifting upwards will be eventually extracted into the gaseous phase thanks to a stronger electric field ($\sim$2\,kV/cm in the liquid) provided by the extraction grid placed 5\,mm below the liquid surface. Once in gas argon, electrons will enter the LEMs, where they undergo charge amplification. LEMs are constructed by drilling submillimiter diameter holes, spaced by a fraction of a mm in a thin printed circuit board, followed by Cu-etching of the hole's rim~\cite{Breskin2013}. An electrical potential is applied between the conductive plates, creating a strong electric field in the holes ($\sim$30\,kV/cm, which is below the breakdown voltage of GAr). The drifting electrons are collimated in the holes where they accelerate and create a cascade of secondary electrons, called ``Townsend avalanche''.
The charge is collected by a multilayer PCB anode~\cite{Cantini2014}. 
The anode is designed to provide two fully symmetric collection planes, meaning the signal is unipolar only. For each plane a gain of 10 for the charge extracted is foreseen during operations. 

Given we have now outlined both single- and dual-phase LAr TPC technology and presented the design of both detectors, we may summarise their differences, identifying advantages and disadvantages. 

\section{Final remarks}
Regarding the ProtoDUNE-SP one clear advantage is that there is no need to adjust the liquid level; instead, for the ProtoDUNE-DP one needs to precisely position the CRP above the liquid surface and keep surface instabilities to a minimum. 

Since  the SP is segmented into two 3.6\,m wide detectors, the high-voltage feedthrough needs to deliver roughly half of the DP voltage for the same electric drift field. On the other end, the longer drift distance is an elegant feature of the ProtoDUNE-DP, as it allows to construct the TPC with minimum amount of channels and to have one single active volume. 

Another advantage of the SP versus the DP is that while in the ProtoDUNE-DP the high-voltage feedthrough needs to go from the top to the cryostat down to the cathode, which adds some mechanical challenges, for the ProtoDUNE-SP the high-voltage feedthrough only needs to penetrate the insulation layer ($\sim$2\,m) in order to touch the cathode plane.  

On the other end, the DP technology has the huge advantage of relying on two symmetric collection planes, which provide a clearer signal than the induction planes of the SP (smaller, bipolar signals are also harder to simulate and reconstruct than unipolar). Finally, the large signal to noise ratio permits operation at higher granularity (3\,mm versus the 5\,mm on the SP), besides making the detector less sensitive to the grounding scheme. 

\bibliography{refs}

\end{document}